\documentclass[%
reprint,
 groupedaddress,
 twocolumn,
 amsmath,amssymb,
 aps,
 prl,
]{revtex4-1}

\usepackage{graphicx}
\usepackage{dcolumn}
\usepackage{bm}
\usepackage{braket} 
\usepackage[colorlinks=true, linkcolor=cyan, urlcolor=blue]{hyperref}
\usepackage[export]{adjustbox} 
\usepackage{tabularx}
\usepackage{todonotes}

\usepackage[flushleft]{threeparttable}
\usepackage{tikz}
\usetikzlibrary{calc}

\begin{document}

\title{Effective spin-orbit models using correlated first-principles wave functions}

\author{Yueqing Chang}
\author{Lucas K. Wagner}
\affiliation{Department of Physics, University of Illinois at Urbana-Champaign, Urbana, Illinois 61801, USA }

\begin{abstract}
  Diffusion Monte Carlo is one of the most accurate scalable many-body methods for solid state systems. 
  However, to date, spin-orbit interactions have not been incorporated into these calcualtions at a first-principles level; only having been applied to small systems.
In this technique, we use explicitly correlated first-principles quantum Monte Carlo calculations to derive an effective spin-orbit model Hamiltonian.
The simplified model Hamiltonian is then solved to obtain the energetics of the system.
To demonstrate this method, benchmark studies are performed in main-group atoms and monolayer tungsten disulfide, where high accuracy is obtained.  
\end{abstract}

\maketitle

The interplay between the electron correlation and relativistic effects may give rise to a plethora of intriguing quantum phases in condensed matter systems, including the proposed axion insulating state, topological Mott insulating state, Weyl semimetal and quantum spin liquid, etc. \cite{pesin2010mott, witczak2014correlated}.
Spin-orbit interactions in materials have been widely investigated in the weakly interacting regime using mean-field theory, which has resulted in the prediction of topological phases \cite{kane2005quantum, bernevig2006quantum, qi2011topological}. 
The spin-orbit effect can remove orbital degeneracies and modify the electronic and magnetic structures of materials; it thus finds applications in spintronic devices \cite{vzutic2004spintronics}.
In the strong-to-intermediate correlation regime, spin-orbit interaction is proposed to play important roles in unconventional Mott insulating and topological semimetal states \cite{kim2009phase, wan2011topological}, quantum spin liquid \cite{alpichshev2015confinement,law20171t}, axion insulating state \cite{go2012correlation} and multipolar orders \cite{chen2010exotic, chen2011spin}.

It is a major challenge to describe both electron correlation and spin-orbit effects in materials. 
Many-body wave function techniques are the most direct way of accessing electron correlations with high accuracy.
In particular, quantum Monte Carlo techniques have seen success \cite{wagner2015ground, yu2015towards, mitra2015many, yu2017fixed} in computing the properties of materials with strong correlation. 
However, to our knowledge, there are no first-principles quantum Monte Carlo calculations on bulk materials that also include spin-orbit effects, previous to this research.

Spin-orbit effects combined with electron correlation are commonly considered for small molecular systems. 
Expansion of the spin-orbit interaction operator in terms of configuration state functions suffer from a slow convergence of the electron correlation contribution \cite{hess2003relativistic}.
As for methods based on stochastic sampling procedures, a recent work employs both an auxiliary continuous parameter to represent the electron spin in the wave function and a modified sampling process in the quantum Monte Carlo (QMC) calculation \cite{ambrosetti2012variational, melton2016spin, melton2016quantum}. 
This method is shown to obtain accurate results for atoms and few-atom molecules, two-dimensional homogeneous electron gas, quantum wells, and circular quantum dots with Rashba interaction \cite{ambrosetti2009quantum, ambrosetti2011spin, ambrosetti2011quantum}.
However, since the spin-orbit energy and the total Coulomb energy are treated on the same footing (a positive feature for some reasons), the computational cost needed to resolve the tiny energy differences among different spin configurations can be very high for solid state systems.

In this paper, we demonstrate a method for treating spin-orbit interactions using quantum Monte Carlo in a scalable and accurate way through rigorous model Hamiltonians. 
We implement spin-orbit interaction calculations using samples of static-spin wave functions, then downfold this set of first-principles data to a model Hamiltonian.
On this manifold, the computational cost needed to resolve different spin configurations is greatly reduced compared to dynamical spins in the calculation, very similar to that needed in a standard fixed-node diffusion Monte Carlo (FN-DMC) calculation. 
To demonstrate our method, we perform benchmark studies in small atomic systems and a solid state system.
Using this technique, we calculate the ground state fine-structure splittings induced by spin-orbit interaction for main-group atoms.
Our calculations generate values agree with the experiments for most atoms, with errors attributed mainly to the lack of $jj$ coupling in the calculation.
To show that this method can be easily generalized to realistic materials, we compute the band splitting at $K$ point for monolayer tungsten disulfide (WS$_2$), yielding a value of $0.39(2)$ eV, as compared to the photoluminescence measurement of $0.4$ eV \cite{zhao2012evolution}.

\paragraph{Effective spin-orbit Hamiltonian.}
Our method is quite general and can treat effective interactions as well as multiple bands \cite{changlani2015density, zheng2018from}.
However, for this paper, we will focus on the case of electronic states that are degenerate in the absence of spin-orbit interaction; for example the six occupations of a single electron on a $p$ manifold. 
Consider many-body wave function $\ket{\Psi_N}$ within that degenerate subspace. 
Then the expectation value of the energy is given by 
\begin{equation}
\braket{\Psi_n | \hat{H}_{\text{rel}} | \Psi_n} = E_0 + E_{\text{SOI}}[\Psi_n],
\label{eqn:soi_split}
\end{equation}
where we evaluate $E_{\text{SOI}}$ by subtracting an averaged-relativistic effective core potential (ARECP) from the relativistic effective core potential (RECP) \cite{melton2016quantum}:
\begin{equation}
\begin{split}
\widehat{W}_{\text{SOI}} = &\sum_{i=1}^{N_{\text{elec}}}\sum_{I=1}^{N_{\text{ions}}}\sum^{L-1}_{l=0} \frac{2}{2l+1}
\left(  \hat{v}^{iI}_{l,j=l+\frac{1}{2}} - \hat{v}^{iI}_{l,j=l-\frac{1}{2}}\right)\\
&\times
\sum^l_{m_l=-l} \sum^l_{m'_l = -l}
\ket{lm_l}\bra{lm_l}\hat{\mathbf{l}}^i\cdot \hat{\mathbf{s}}^i
\ket{lm'_l}\bra{lm'_l}.
\end{split}
\label{eqn:wsoi}
\end{equation}
where operators $\hat{v}^{iI}_{l,j=l\pm\frac{1}{2}}$ are the spin-up/down radial parts of the two-component RECP between the $i$th electron and the $I$th ion, respectively.
The spin-orbit energy in Eq.~\ref{eqn:soi_split} is given by $E_\text{SOI}[\Psi_n]=\bra{\Psi_n}\widehat{W}_{\text{SOI}}\ket{\Psi_n}$.

We take the effective Hamiltonian of the degenerate manifold to be
\begin{equation}
\widehat{H}^{\text{eff}}_{\text{rel}} = E_0 + \lambda \sum_{i=1}^{N_{\text{elec}}} \sum_{I=1}^{N_{\text{ions}}}\widehat{\mathbf{L}}_{iI} \cdot\widehat{\mathbf{S}}_i,
\label{eqn:eff_soi}
\end{equation}
where the operator $\widehat{\mathbf{L}}_{iI} = \left(\mathbf{r}_i-\mathbf{r}_I \right)\times \frac{1}{\mathrm{i}} \nabla_i$ is the orbital angular momentum of the $i$th electron relative to $I$th ion, and $\widehat{\mathbf{S}}_i$ is the spin angular momentum operator acting on the $i$th electron.
\begin{figure}
\centering
\includegraphics{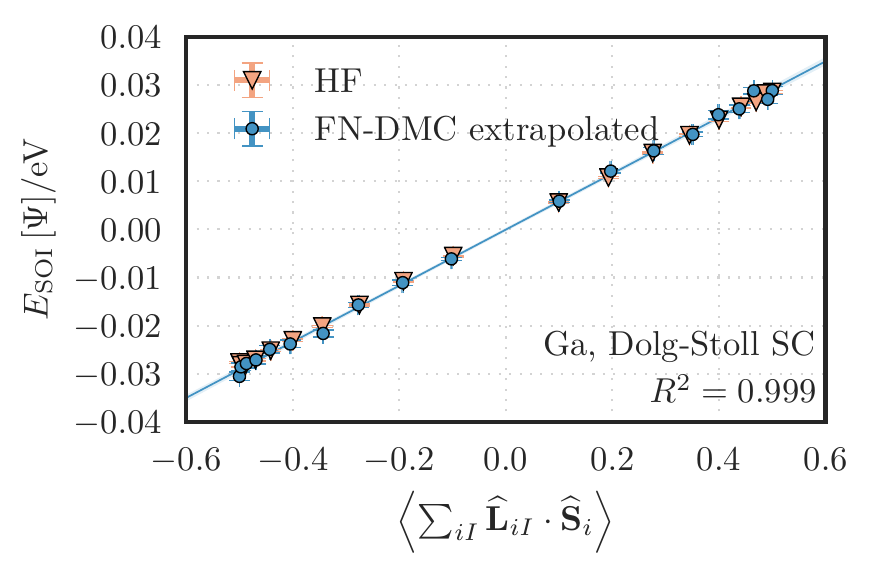}
\vspace{-0.6cm}
\caption{Calculated spin-orbit interaction energies versus SOI descriptors $\braket{\sum_{iI} \widehat{\mathbf{L}}_{iI}\cdot \widehat{\mathbf{S}}_i}$ for 20 wave function samples for Ga atom. 
The Dolg-Stoll small-core ECP is used.
The red triangles (HF) denote results calculated using VMC with Slater-determinants.
The blue circles (FN-DMC extrapolated) are the FN-DMC extrapolated estimators.}
\label{fig:fig1}
\end{figure}

Using the results proved in \cite{changlani2015density,zheng2018from}, the effective $\lambda$ can be estimated using the first-principles wave functions $\ket{\Psi_n}$ by writing
\begin{equation}
\braket{\Psi_n | \hat{H}_{\text{rel}} | \Psi_n} = E_0 + \lambda \braket{ \Psi_n | \sum_{i=1}^{N_{\text{elec}}} \sum_{I=1}^{N_{\text{ions}}} \widehat{\mathbf{L}}_{iI} \cdot\widehat{\mathbf{S}}_i | \Psi_n }.
\label{eqn:fit}
\end{equation}
Using Eq.~\ref{eqn:soi_split}, we thus have 
\begin{equation}
E_\text{SOI}[\Psi_n] = 	\lambda \braket{ \Psi_n | \sum_{i=1}^{N_{\text{elec}}} \sum_{I=1}^{N_{\text{ions}}} \widehat{\mathbf{L}}_{iI} \cdot\widehat{\mathbf{S}}_i | \Psi_n }
\label{eqn:e_soi}
\end{equation}
This relationship is true for any wave function $\ket{\Psi_n}$ sampled from the degenerate manifold if the spin-orbit model is valid.
Since both $E_\text{SOI}[\Psi_n]$ and $\braket{ \Psi_n | \sum_{iI} \widehat{\mathbf{L}}_{iI} \cdot\widehat{\mathbf{S}}_i | \Psi_n }$ can be computed using first-principles wave functions, the coefficient $\lambda$ can be obtained by fitting to the linear relationship between these two values.
For periodic systems, the effective Hamiltonian at momentum $\mathbf{k}$ is given by $\widehat{H}_{\mathbf{k},\text{SOI}}^{\text{eff}} = \sum_{iI}\lambda_{I}(\mathbf{k}) \widehat{\mathbf{L}}_{iI}\cdot \widehat{\mathbf{S}}_{i}$ (for the monolayer WS$_2$, the spin-orbit interaction is mostly contributed by W atoms). 
This method can be easily extended to the case where the wave functions do not belong to a degenerate manifold by including parameters that represent orbital energies, interaction terms, and so on. 
This has been performed for non-spin-orbit-coupled systems \cite{zheng2018from}.

\paragraph{Details of the calculations.} 
For a condensed matter system, we first construct single Slater determinants from the single-particle orbitals computed using Hartree-Fock or density functional theory.
These Slater determinants are sampled such that they differ only by single-particle orbitals that have the same scalar-relativistic energy.
We construct the following Slater-Jastrow trial wave functions,
\begin{equation}
\Psi_{\text{T}} = e^{U} D^{\uparrow} D^{\downarrow},
\label{eqn:wavefunction}
\end{equation}
where the spins are collinear, chosen to be along the $z$ axis.
Determinants $D^{\uparrow}$ and $D^{\downarrow}$ are constructed from occupying the lowest energy spin-up and spin-down single-particle orbitals, respectively.
To incorporate the electron correlation explicitly, the trial wave functions are constructed by multiplying these Slater determinants by Jastrow factors $U$s which include up to three-body interactions (see \cite{wagner2009qwalk}). 
The Jastrow factors are then optimized by minimizing the variance of the trial scalar-relativistic energy.
Finally, we apply FN-DMC to project out the lowest energy wave function that has the same nodal structure as the corresponding trial wave functions.
The scalar-relativistic Hamiltonian we use to evolve the trial wave function in FN-DMC is given by expressing the Coulomb potential using the sum of electron-electron interaction among the valence electrons and the averaged-relativistic effective core potential (ARECP).
$T$-moves are used to evaluate the nonlocal effective core potentials during the imaginary-time evolution of the trial wave functions \cite{casula2006beyond}. 
In order to correct the time-step error introduced by Trotter product, we extrapolate the result to zero-time-step limit using a method based on the Bayesian probability distribution described in \cite{wagner2010quantum}.
All VMC and FN-DMC calculations are performed using the open-source quantum Monte Carlo code QWalk \cite{wagner2009qwalk}.

Because the \textit{ab initio} spin-orbit interaction Hamiltonian (Eq.~\ref{eqn:eff_soi}) does not commute with the scalar-relativistic Hamiltonian, we correct the mixed estimator error up to first order by computing the extrapolated estimators given in \cite{foulkes2001quantum}.
We compute the extrapolated estimator values of the spin-orbit interaction energies and the corresponding descriptor $\braket{\sum_{iI} \widehat{\mathbf{L}}_{iI} \cdot \widehat{\mathbf{S}}_{i}}$ for several fixed-node wave functions, then find the coefficient $\lambda$ using linear regression.
This procedure maps a set of values calculated from first-principles wave functions to a model Hamiltonian Eq.~\ref{eqn:eff_soi}.
Such procedures can be generalized to expectation values calculated using dynamic-spin wave functions, but for the brenchmarking systems considered here, static-spin wave functions are sufficient.  

For the atomic systems, we use GAMESS\cite{schmidt1993general, dykstra_chapter_2005} to generate the single-particle orbitals for the cations and anions (Ga$^+$, In$^+$, Ge$^{2+}$, Sn$^{2+}$, Br$^-$ and I$^-$), such that the $np_x$, $np_y$ and $np_z$ orbitals have exactly the same energies for each ion.
Thus, we can reduce the orbital relaxation effect introduced by performing restricted open-shell Hartree-Fock calculations directly on atomic systems, 
Then, we use these ionic orbitals to construct the Slater determinants for the corresponding atoms.
We perform the calculations using the Dolg-Stoll and the Trail-Needs ECPs and the available basis sets provided in \cite{metz2000small-Pb, stoll2002relativistic, peterson2003systematically, peterson2006spectroscopic, trail2005smooth, trail2005norm}.
The Dolg-Stoll ECPs include relativistic effects implicitly by adjusting the orbitals at the multiconfiguration Dirac-Hartree-Fock level and including the Breit interaction \cite{metz2000small-Pb, stoll2002relativistic, peterson2003systematically, peterson2006spectroscopic}. 
We also compare the large-core and small-core Dolg-Stoll ECPs, which exclude or include respectively the semi-core $(n-1)spd$ shells in the valence shell. 
The Trail-Needs ECPs are large-core only for the elements considered here, and are finite at the origin, thus are particularly suitable for QMC calculations \cite{trail2005smooth, trail2005norm}.

For the WS$_2$ calculations, we use CRYSTAL17 \cite{dovesi2018quantum, dovesi2017manual} to perform mean-field calculations for monolayer WS$_2$, with the lattice constant $a=3.23 \text{ \AA}$, and the spacing between two S atoms in the same unit cell $d=3.19 \text{ \AA}$.
The molecular orbitals are calculated for the primitive cell by solving the Kohn-Sham equation, where the exchange-correlation interactions are included within the generalized gradient approximation in the Perdew-Burker-Ernzerhof form \cite{perdew1996generalized}.
We use a converged $12\times 12$ shrinking factor for reciprocal lattice vectors.
The Dolg-Stoll averaged-relativistic ECP (60 core electrons for W and 10 core electrons for S) is used \cite{figgen2009energy, igel1988pseudopotentials}.
A truncated Gaussian basis set including up to triple $\zeta$ functions is used to expand the molecular orbitals for W \cite{figgen2009energy}.
For S, we use the starting basis set provided in reference \cite{igel1988pseudopotentials}.

\paragraph{Benchmark studies.}
\begin{figure}
 \flushleft \footnotesize{(a).}\\
\includegraphics{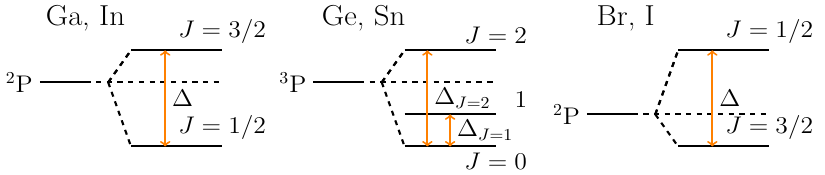}
 \flushleft \footnotesize{(b).}\\
\includegraphics{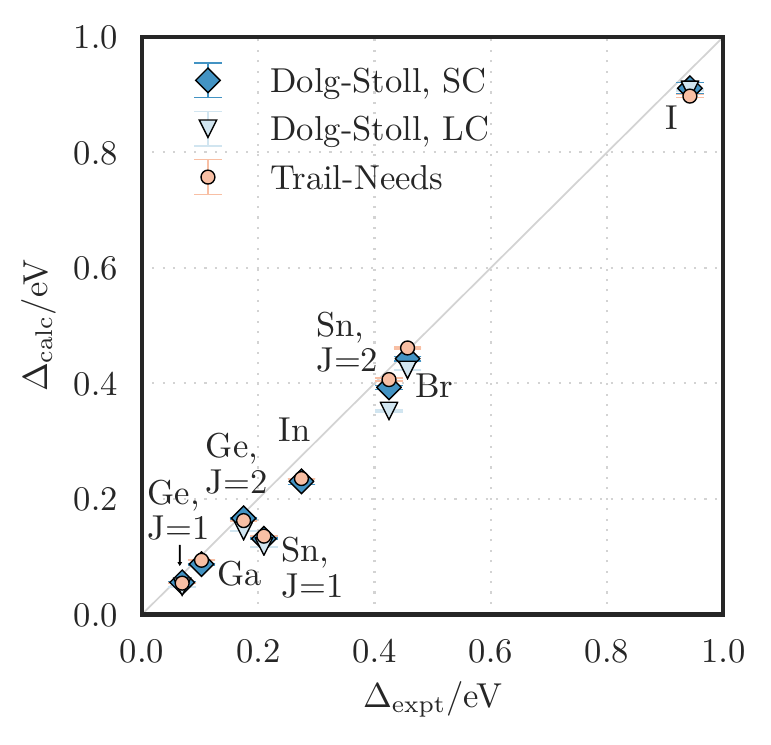}
\caption{
 (a). Spin-orbit splittings of the ground state electron configurations of Ga, In, Ge, Sn, Br, and I atoms;
(b). Fixed-node DMC extrapolated estimators (the mixed estimators are extrapolated to the zero time-step limit) versus the experimental values of the fine-structure splittings for the chosen main-group atoms. 
Different colors and shape of markers denote calculated values using different ECPs.}
\label{fig:fig2}
\end{figure}
  
Taking the Ga atom as an example, the ground-state electron configuration without spin-orbit interaction is [Ar]$3d^{10}4s^24p^1$ with $L=1$, $S=\frac{1}{2}$ ($^2P$, 6-fold degenerate).
The spin-orbit interaction splits the 6-fold degenerate state into two manifolds, the $J=\frac{3}{2}$ quartet and the $J=\frac{1}{2}$ doublet (see Fig.~\ref{fig:fig2}, panel (a)). 
In the Ga atom, the sampling of wave functions on a sub-Hilbert space with the lowest scalar-relativistic energy is performed by projecting out the fixed-node wave functions from the trial Slater-Jastrow wave functions, whose Slater parts are linear superpositions of the Slater determinants including $4p_x$, $4p_y$ and $4p_z$ orbitals.
In our calculations, we generate 20 samples of such trial wave functions with different SOI descriptors values $\braket{\sum_{iI} \widehat{\mathbf{L}}_{iI} \cdot \widehat{\mathbf{S}}_i}$ ranging from $-0.5$ to $0.5$ in atomic units, when taking the dimensionless $S_z$ for spin-up/down electron to be $\pm 1/2$. 
Within this context, the spin-orbit splitting $\Delta$ of Ga atom ground-state configuration is given by $\Delta = 3/2 \lambda$.

In Fig.~\ref{fig:fig1}, we plot the SOI energies $E_{\text{SOI}}$ versus the descriptors $\braket{\sum_{iI} \widehat{\mathbf{L}}_{iI} \cdot \widehat{\mathbf{S}}_i}$ calculated on 20 Slater determinants using variational Monte Carlo (VMC) (labeled by the red triangles), and on the corresponding fixed-node wave functions for Ga atom (labeled by the blue circles). 
The Dolg-Stoll small-core (10 core electrons) and (12$s^{ }$12$p^{ }$9$d^{ }$)/[6$s^{ }$6$p^{ }$4$d^{ }$] basis set are used \cite{metz2000small-Pb}. 
There is a linear relationship between these two quantities, as required by Eq.~\ref{eqn:fit}.

To demonstrate the accuracy of the technique, we compute the spin-orbit splittings of the ground state configurations of the following main-group atoms: Ga, In, Ge, Sn, Br, and I.
The In, Br and I atom have similar spin-orbit splitting pattern as Ga, while the ground state splittings of Ge and Sn atoms $\Delta_{J=1}$ and $\Delta_{J=2}$ satisfy $\Delta_{J=1} = 1/2 \left|\lambda\right|$ and $\Delta_{J=2} = 3/2 \left|\lambda\right|$ (see Fig.~\ref{fig:fig2} (a)).

To check if there is dependence on the ECP, we perform the same calculations using the Dolg-Stoll and the Trail-Needs ECPs \cite{metz2000small-Pb, stoll2002relativistic, peterson2003systematically, peterson2006spectroscopic, trail2005smooth, trail2005norm}.
Fig.~\ref{fig:fig2} (b) shows the FN-DMC extrapolated estimator values of the SOI splittings of the ground state configurations, versus the experimentally determined fine-structure splittings for the chosen main-group atoms (extracted from NIST atomic spectra database \cite{ralchenko2008nist}).
The light gray line indicates an equality between the calculated values and the experimentally determined values.
From the figure, we can see that our method yields better agreement with the experiments when using the Dolg-Stoll small-core ECP and the Trail-Needs ECP, compared with that when using the Dolg-Stoll large-core ECP.

We attribute the discrepancy between the calculated SOI splitting and the experimentally determined fine-structure splitting of the Sn atom $J=1$ state to the interplay of the $jj$-coupling and the $LS$-coupling (the spin-orbit interaction).
Our model Hamiltonian Eq.~\ref{eqn:eff_soi} only includes the spin-orbit interaction, thus the Land\'e interval rule follows, i.e., $\Delta_{J=2}/\Delta_{J=1} = 3/1$.
However, due to the non-negligible effect of the $jj$-coupling, the experimentally observed fine-structure splittings of Sn atom is not entirely induced by spin-orbit interaction of electrons. 
As a consequence, Land\'e interval rule does not apply, i.e., $\Delta_{J=2}/\Delta_{J=1} \neq 3/1$.

\begin{figure}
   \centering
   \includegraphics{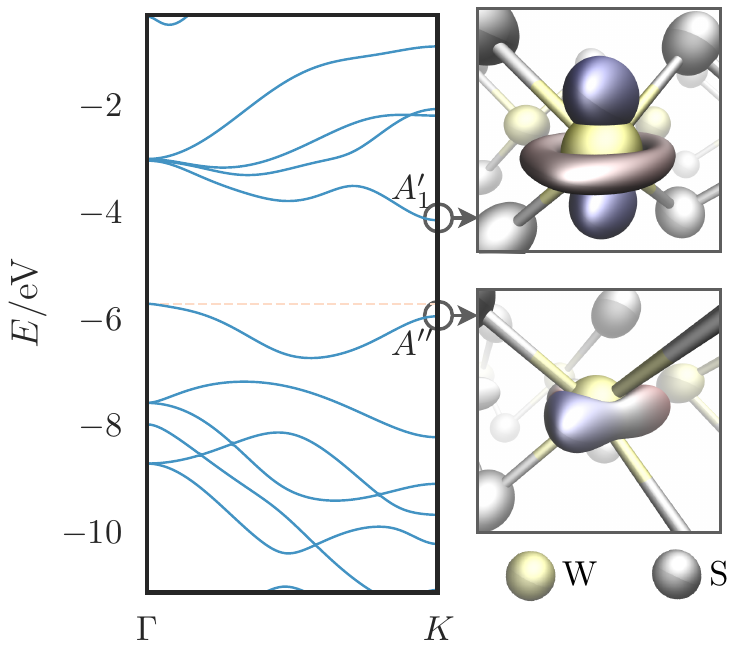}
   \caption{Band structure of monolayer WS$_2$ calculated on the unit cell, using the Perdew-Burke-Ernzerhof exchange-correlation functional \cite{perdew1996generalized} without spin-orbit interaction. 
The red dashed line represents the Fermi level.
The conduction band minimum and valence band maximum at $K$ are labeled by $A'_1$ and $A''$.
The corresponding single-particle orbitals are plotted on the right using the Visual Molecular Dynamics software \cite{HUMP96}, with orbital shape derived from the isosurface of the magnitude, colored by the phase.
  }
     \label{fig:fig3}
\end{figure}

In order to demonstrate that our model can be easily generalized to larger scale systems, we perform a benchmark study in monolayer WS$_2$. 
Due to the presence of tungsten atoms, the strong spin-orbit interaction splits the valence band maximum at $K$. 
We apply our method to compute this band splitting.
We construct trial wave functions from superpositions of four Slater determinants multiplied by three-body Jastrow factors. 
These Slater determinants are constructed by promoting one spin-up/down electron from the two-fold degenerate valence band maximum ($A''$) to the conduction band minimum ($A'_1$) at $K$, such that they have degenerate scalar-relativistic energy (see Fig.~\ref{fig:fig3}).

\begin{figure}
\centering
\includegraphics[width=0.5\textwidth]{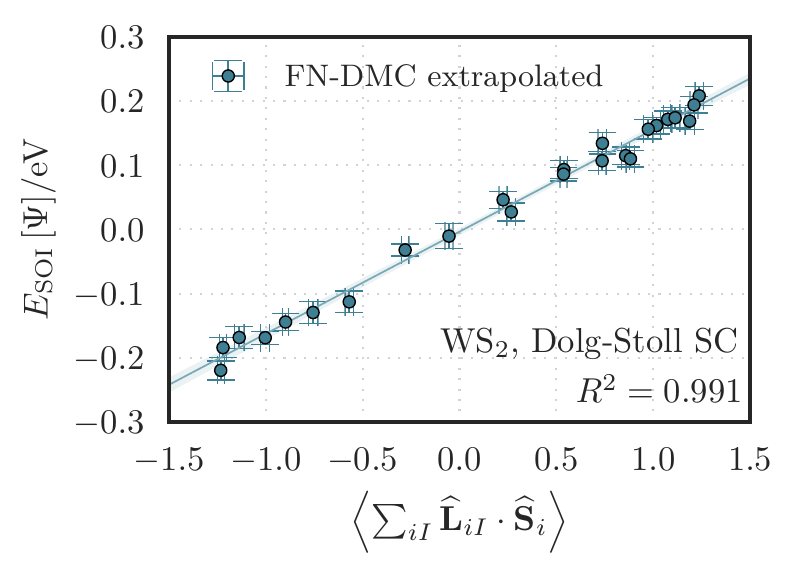}
\vspace{-0.8cm}
\caption{Calculated spin-orbit interaction energies versus SOI descriptors $\braket{\sum_{iI} \widehat{\mathbf{L}}_{iI} \cdot \widehat{\mathbf{S}}_i }$ for 24 fixed-node wave functions constructed on a manifold that has degenerate scalar-relativistic energy at $K$ for monolayer WS$_2$. 
The circles are the FN-DMC extrapolated estimators, which are fitted to the solid line. }
\label{fig:fig4}
\end{figure}


Fig.~\ref{fig:fig4} shows the FN-DMC extrapolated SOI energies versus the SOI descriptors calculated using samples of many-body wave functions of monolayer WS$_2$ at $K$ point.
The slope of the fitted line is the spin-orbit interaction strength $\lambda$ in our linear model. 
The calculated spin-orbit splitting $\Delta$ for the valence band maximum at $K$ is $0.39(2)$ eV. 
This is statistically in agreement with the band splitting $0.43$ eV calculated using PBE exchange-correlation functional reported in reference \cite{kang2013band}.
It also agrees with the reported band splitting $0.4$ eV, determined by measuring the differential reflectance spectra of exfoliated single layer 2H-WS$_2$ \cite{zhao2012evolution}.
This demonstrates that our method can be easily generalized to perform spin-orbit interaction calculations in solid state systems.


\paragraph{Conclusion.}
  
We have demonstrated a scalable first-principles quantum Monte Carlo technique for extended systems; this will allow future studies to blend treatment of electron correlation and spin-orbit interactions using QMC calculations. 
The method is based on deriving an effective Hamiltonian for the spin-orbit interaction.
In this method, the major computational cost is contributed by the evaluation of spin-orbit interaction energy on the FN-DMC wave functions, less than twice that of a standard FN-DMC calculation.
One can include also electron-electron interactions in the effective Hamiltonian. 

We demonstrated this technique in atomic systems and monolayer WS$_2$.
For the main-group atoms, we compute the spin-orbit splittings of the ground state configurations and obtain results in agreement with the experimentally determined fine-structure splittings.
For the monolayer WS$_2$, we compute the band splitting induced by spin-orbit interaction at $K$.
Our first-principles result agrees with previous calculations and experiments, thus demonstrating the ability of this method to be generalized to larger scale systems.

A major promise of this technique is that one can treat electron-electron interactions, spin-orbit effects, and one-body terms in effective Hamiltonians all on the same footing. 
As mentioned before, the cost is similar to a standard FN-DMC calculation, which will allow it to be applied to realistic models of materials. 
We envision this opening a new frontier for modeling spin-orbit effects in correlated materials.

\begin{acknowledgements}
This work was funded by the grant DOE FG02-12ER46875 (SciDAC). The authors want to thank Cody Melton for precious discussions. 
\end{acknowledgements}

\bibliography{cite}

\end{document}